\documentclass[runningheads]{llncs}
\usepackage{graphicx}
\usepackage{mathtools}
\usepackage{amsmath}
\usepackage{amssymb} 
\usepackage{color}
\usepackage{mathabx}
\usepackage{booktabs}
\usepackage[width=122mm,left=12mm,paperwidth=146mm,height=193mm,top=12mm,paperheight=217mm]{geometry}

\DeclareMathOperator*{\argmax}{argmax}

\begin{document}

\renewcommand\floatpagefraction{.9}
\renewcommand\topfraction{.9}
\renewcommand\bottomfraction{.9}
\renewcommand\textfraction{.1}
\setcounter{totalnumber}{50}
\setcounter{topnumber}{50}
\setcounter{bottomnumber}{50}
\addtolength{\itemsep}{-0.05in}
\addtolength{\topsep}{-0.07in}
\addtolength{\textfloatsep}{-0.05in}
\addtolength{\intextsep}{-0.05in}
\addtolength{\partopsep}{-0.03in}
\addtolength{\parskip}{-0.02in}

\pagestyle{headings}
\mainmatter
\def\ECCV18SubNumber{***}  

\title{Understanding the Gist of Images - Ranking of Concepts for Multimedia Indexing} 

\titlerunning{Understanding the Gist of Images}

\author{Lydia Weiland\inst{1} \and Simone Paolo Ponzetto\inst{1} \and Wolfgang Effelsberg\inst{1} \and Laura Dietz\inst{2}}
\authorrunning{L. Weiland, S. P. Ponzetto, W. Effelsberg, and L. Dietz}
%
%
%
\institute{University of Mannheim, Germany \and University of New Hampshire, Durham, New Hampshire, USA}

\maketitle

\begin{abstract}
Nowadays, where multimedia data is continuously generated, stored, and distributed, multimedia indexing, with its purpose of grouping similar data, becomes more important than ever. 
Understanding the gist (=message) of multimedia instances is framed in related work as a ranking of concepts from a knowledge base, i.e., Wikipedia. 
We cast the task of multimedia indexing as a gist understanding problem.  
Our pipeline benefits from external knowledge and two subsequent learning-to-rank (l2r) settings. 
The first l2r produces a ranking of concepts representing the respective multimedia instance. 
The second l2r produces a mapping between the concept representation of an instance and the targeted class topic(s) for the multimedia indexing task. 
The evaluation on an established big size corpus (MIRFlickr25k, with 25,000 images), shows that multimedia indexing benefits from understanding the gist. 
Finally, with a MAP of 61.42, it can be shown that the multimedia indexing task benefits from understanding the gist. 
Thus, the presented end-to-end setting outperforms DBM and competes with Hashing-based methods.   
\end{abstract}

\section{Introduction}\label{UseCase:Intro}
Conveying meaning by the use of different modalities is probably as old as human mankind. Nowadays, where everyone takes pictures constantly, storage is cheap, and several platforms allow for the distribution, sharing, and modification of multimodal data, the need of adequately representing the content of such multimodal data is more important than ever. 
Research on multimedia data, independent of the goal - i.e., whether it is modality generation, modality retrieval, or representation of the modalities - have shown 
to perform better with a joint representation of the different modalities. 
Thus, multimedia researchers motivate the joint modeling by observing ("people often caption an image to say things that may not be obvious from the image itself, such as the name of the person, place, or a particular object in the picture."~\cite[p. 2950]{DBM2012}). 

In this work, we focus on multimedia representation. Thus, we make use of the MIR Flickr dataset~\cite{Huiskes2008}, which provides - besides the images and their affiliated tags - a gold standard multi-label annotation, where the labels represent semantic class topics, such as 'sky' or 'people'. 
A lot of research has been conducted providing a detailed analysis on the MIR Flickr dataset, these works can be mainly grouped into deep learning, auto-encoder, hashing-based and other approaches. 
To the best of our knowledge, none of the previous works have made benefit from knowledge bases, such as Wikipedia. 

\medskip \noindent \textbf{Task:} Given an image and its tags, represented by concepts from a knowledge base, create a mapping between the representing concepts and the target classes.    

\medskip \noindent To represent images, we adapt, modify, and expand the work of~\cite{weiland16a}. Their task is to understand the \textit{gist} (equivalent to meaning or message) of an image-caption pair. Their idea is to create query-specific knowledge graphs, where both, an image-caption pair and the gist are represented by concepts from the knowledgebase. There are two modifications to their methodology: First, we need to create a mapping between the ranked list of concepts and the target class topic(s) to conduct the multimedia indexing. Second, we encode a graph creation strategy, which has shown to be useful for text-lexical understanding~\cite{NavigliLapata:10,ponzetto10b}. This modified gist pipeline is an end-to-end setting, in that the method relies on out-of-the-box object detectors and a model that has been pre-trained on a dataset with the purpose of gist understanding. 
In an experimental evaluation, we study the per class and the overall performance of the modified gist pipeline. We compare these results with shallow, deep, hashing, and auto-encoder approaches. We find that not only representing the pair as concepts, but the expansion steps to collect semantically related concepts, has a positive impact to the overall performance. Overall, we confirm that the combination of tags and a deep learning based automatic object detection - even though the latter encodes noise - achieves better performance than single modality representations. 

\section{Related Work}\label{sec:rw}


 
Representing multimedia data is often approached as a classification task, where the goal is to assign the multimedia instances to one or several classes best representing the content of the instance. 
In context of multimedia data, which consists of image and text, i.e., the best representing class(es) is given as the class that best describes the most salient object(s) in the image~\cite{Huiskes2008}.
The originators of the MIR Flickr benchmarking dataset presented two different approaches for classification of the data. An SVM and Linear Discriminant Analysis (LDA) based classification.
Both approaches are trained with different feature sets: one using uni-modal features (low-level features from the vision domain) and one combining the low level visual features with the textual tags as features~\cite{Huiskes10}.
There are several successors using various types of approaches, however, none benefit from query-specific knowledge graphs. ~\cite{Hare2010} presents a high-dimensional vector space representations based on a cross-lingual latent semantic indexing. 
Others make use of deep-learning approaches, e.g., multimodal deep boltzman machine (DBM)~\cite{DBM2012}, regularized deep neural network (RE-DNN)~\cite{Wang2016}, deep hashing~\cite{Wu2017}. 

\section{Methodology}\label{sec:method}

 \begin{figure*}[t]
\begin{center}
 \includegraphics[width=\textwidth, height=100.0pt]{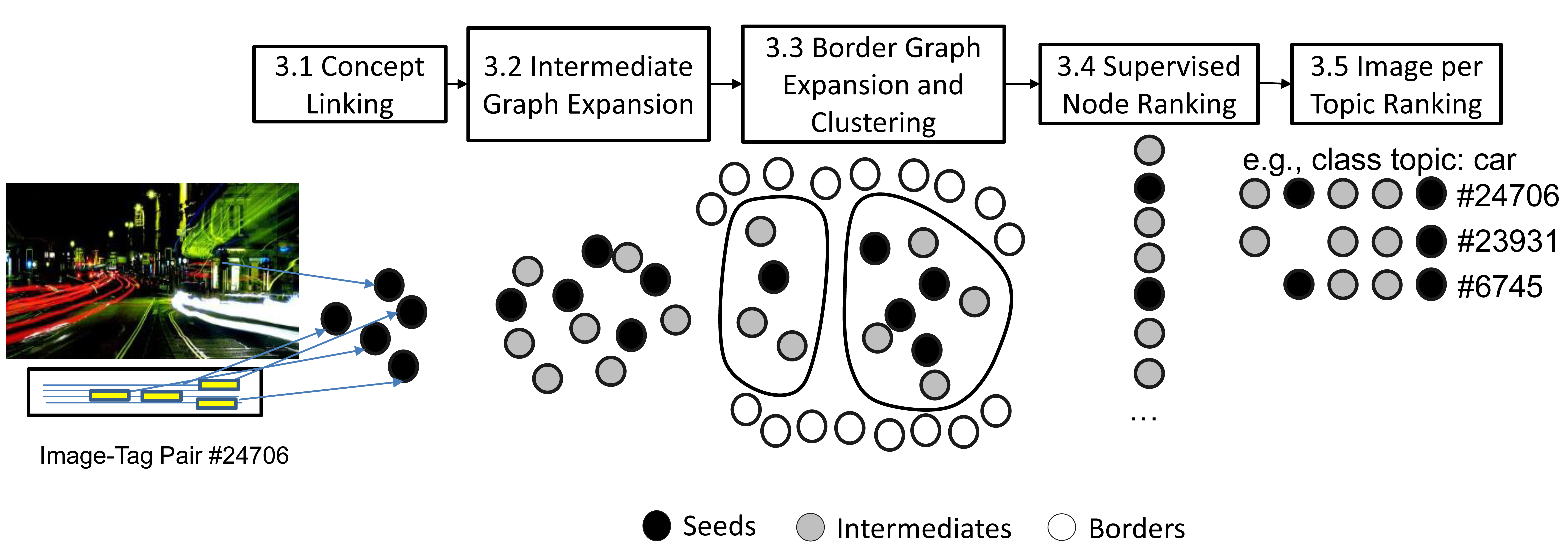}
 \caption[Modified gist pipeline]{The twice-learning-to-rank pipeline for the multimedia indexing task. Image-text pairs are represented as concepts from a knowledge base (Image \#24706: Flickr/clare savory, CC BY 2.0).}
  \label{fig:Modpipeline}
\end{center}
\end{figure*}
Since we base our method on the pipeline introduced by~\cite{weiland16a}, we focus on our novel changes (cf. Fig.~\ref{fig:Modpipeline}). 
However, we start with an overview of the complete pipeline and indicate - where required - the modifications compared to the original pipeline.

The main idea of the approach is to represent the images and their affiliated text by concepts from a knowledge base and conduct a query-specific knowledge graph generation, which in the case of~\cite{weiland16a} represents the gist and which in our case is used to create a mapping between query image and target class. 
In the knowledge base representation of Wikipedia we are using, article pages and categories are the concepts.
Besides the representation of Wikipedia as a knowledge base, we benefit from a representation as a knowledge graph, where categories and articles are nodes in the graph. The article redirects and the category links are the edges in the graph.


\medskip \noindent \textbf{Seed Node Linking}
As we aim at an end-to-end setting, we benefit from a state-of-the-art image processing API, such as Microsoft Cognitive Services~\footnote{Note: By End of 2017 the API is integrated into Microsoft Azure.}, to assign textual labels to objects, attributes, and the scenery in the images, e.g., car, yellow, outdoor, respectively.  
We use the textual object labels as candidates for concept mentions.
Additionally, the Flickr tags provided in the MIRflickr25k benchmarking dataset, serve as candidates for the concept linking. 

Via a string-match based concept linking, the concept mentions from both, the image and the text, are linked to concepts of a knowledge base. 
The resulting set of concepts is what is called the seed nodes, consisting of nodes with origin in the image or the textual Flickr tags, in the following we refer to these concepts as \textbf{$S_{i}$} and \textbf{$S_{t}$}, or to their combination as \textbf{$S_{t, i}$}, respectively.

\medskip \noindent \textbf{Intermediate Graph Creation.}\label{sec:method_inter} Under the assumption that the image and the text convey some specific - sometimes complementary - aspect of a common meaning, the next step is to search for semantic connections between the image and its text. Given the seed nodes from the previous step, we collect all category nodes on shortest paths (up to length 4) between all pairs of seed nodes. Therefore edges representing category links are followed. 
The result is a sub-graph, which consists of seed nodes, the collected category nodes (called intermediate nodes) and the edges connecting the seed and intermediate nodes. In the following we refer to this procedure as the intermediate graph creation and to concepts collected in this step to as \textbf{$I$}.  
\textbf{Example:} One query image has the seed nodes volvo and car, where a node on the intermediate path is Category:Motor vehicles. Another query image has the seed nodes motorcycle and traffic, which are also connected to Category:Motor vehicles. Consequently, the instances which initially did not share a concept, now share a concept. 

By applying Louvain clustering~\cite{blondel2008fast} based on the path based semantic relatedness measure of~\cite{Hulpus2015pathbased}, we address the fact that a query image is of diverse nature, in that it can be assigned to several class topics, consequently, the seed concepts are not about one semantic topic and thus, of different semantic relevance.

\medskip \noindent \textbf{Ranking the Nodes.} To conduct the ranking of concepts best representing the image-text pair, we use 16 different features, e.g., graph connectivity and content based measures, boolean and features based on the article texts of Wikipedia concepts (for further details, please refer to~\cite{weiland16a}). As there is no gold standard about the relevance of the so far collected concepts given (recall: a concept in the knowledge base has an equivalent node in the knowledge graph, but as we are not doing a graph ranking, we switch back the notation of concept), we rely on a pre-trained model using the data and gold standard of~\cite{weiland16a}to rank the concepts.

\medskip \noindent \textbf{Learning-to-rank model (l2r).} Our generated feature vectors serve as input for a list-wise learning-to-rank model~\cite{li2011letor}. In a learning-to-rank setting, the image-text pairs are the set of documents $D$, with $d_i$ as the i-th document (following the notation of ~\cite{li2011letor}).
For each instance the candidate concepts are the set of queries, denoted by $Q$, where $q_i$ is the i-th query.
The j-th image-text pair for a query $q_i$ is represented as a feature vector $x_{i,j} = \phi(q_i, d_{i,j})$ of feature functions $\phi$, such as Betweeness Centrality. 
Finally, $S'={(x_i, y_i)}$ represents the training data for $q_i$, with $y$ denoting the set of labels $\{1, 2,..., 5\}$.
The objective is to find a parameter setting $\widetilde{\phi}$ that maximizes the scoring function: 
$\widetilde{\pi} = \argmax_{\pi_i \in \Pi_i} S(x_i, \pi_i)$.
Specifically, we use RankLib\footnote{https://sourceforge.net/p/lemur/wiki/RankLib}, trained with respect to the target metric Mean-average precision (MAP). For optimization we use coordinate ascent with a linear kernel~\cite{Metzler2007}.
The result is a ranked list of concepts for each image-text pair.

\medskip \noindent \textbf{Image per Topic Ranking.} As the objective is a multi-label classification, the ranked list of concepts need to be mapped to the class topics.
There are several ways of creating such a mapping, such as using lexical and conceptual hierarchies (e.g., WordNet).
The issue with these methods is that each single gist concept needs to be matched to the MIR topics, but tags and images in their single instances are not completely semantically coherent with the topic of a class, e.g., a building will not be matched to the class topic people. However, the combination of the concepts make the meaning, e.g., crowded city street.
Hence, we benefit from an additional learning-to-rank approach, with a feature vector representing the ranked gist concepts of an input image-tag pair with the relevance scores from the first l2r serve as feature for the respective concept. 
Across all ranked concepts we build a lexicon, where the size of the lexicon is the number of dimensions for the feature vectors, which will be created. 
The danger here is that the dimension is higher than the number of training instances, however, we find that the query-specific graph creation leads to a small lexicon.
Thus, finally we formulate: Given the class as a query, the task is to rank images that fit the class highest.

\section{Experimental Evaluation}
In the following, we study the performance of framing the multimedia indexing as a modified gist understanding task conducted on the MIR Flickr dataset, which contains around 25,000 images from Flickr with textual tags. We report the results according to the evaluation measures Mean Average Precision (MAP) and Precision of the first 50 positions of our ranking output (P@50).
Given an image-tags pair the task is to assign at least one label of the 10 general- and 19 subtopics, e.g., sky, clouds, class topics to such a query pair. Some of the topics have less image instances than the other, resulting in an instance per topic range between 116 to 10,373 for the topics baby (relevant) and people (potential), respectively. 

\medskip \noindent \textbf{Multimedia Classification (Multimedia Indexing)}
We provide a comparison to state-of-the-art approaches of the multimedia indexing task compared to the results reported by~\cite{Huiskes10,Chen2016,DBM2012}, relying on their results without re-implementation. 

\begin{table*}
	\caption[MIR Flickr: Classification performance across all topics]{Classification performance of the different gist detection pipeline settings across all topics according to \textbf{MAP} and \textbf{P@50}, compared to the two best methods from~\cite{Huiskes10} (the two low-level settings are discarded here. Original numbers from~\cite{Huiskes10}), the random baseline, the Deep Boltzman Machine (DBM) (numbers and approach of~\cite{DBM2012}), and the 32-bit robust multi-label hashing (RMLH) of~\cite{Chen2016}. 		
\label{tab:miniComp}}	
\begin{scriptsize}
	\begin{center}
		\begin{tabular}{lllllllllllllll} 
			\toprule
		\textbf{Method} & Gist($S_{t, i},I$)& Gist($S_{t, i}$) & Gist($S_{t}$) & &LDA &  SVM  & Random & RMLH, 32-bit & DBM\\ 
		\midrule
	\textbf{MAP} & 61.42  & 45.56 & 35.87 & &49.2 & 47.5 & 12.4 & 65.6 & 52.6\\ 
	\textbf{P@50} & 88.3 & 53.71 & 49.18  & &74.5& 75.8 & 12.4 & - & 79.1\\ 
			\bottomrule
		\end{tabular}
	\end{center}
	\end{scriptsize}
\end{table*}

The extended gist detection pipeline outperforms all of the comparison approaches in terms of precision (88.3, cf. Table~\ref{tab:miniComp}).
Additionally, all but one approach are outperformed according to Mean Average Precision (MAP: 61.42) by the extended gist detection pipeline (cf. Table. Only, the multi-label hashing method performs better (MAP: 65.6). However, there is no result reported for this approach with respect to precision.  

\medskip \noindent \textbf{Comparison of concept candidates}
Finally, we compare the performance of the extended gist pipeline, when considering three different sets of concepts in the gist ranking task: concepts from the seed nodes of the tags (Gist($S_{t}$)), concepts from the seed nodes of the tags and the images (Gist($S_{t, i}$)), and concepts from the seed and the intermediate nodes (Gist($S_{t, i},I$).

The lexicon of Gist($S_{t, i},I$) consists of 1,538 different top-10 concepts. This is an average of 40.5 concepts of representing one class. 
Gist($S_{t, i}$) and Gist($S_{t}$) contain 5,809 and 11,755 concepts, respectively. 
The higher number of concepts for Gist($S_{t}$) underlines the diversity across the concepts using the tags only (cf. Appendix - Seed Node Linking) - which might not be the best characteristic to find semantic similarity. That Gist($S_{t, i},I$) has the smallest number of different concepts, confirms the idea of finding common concepts with the modified gist understanding pipeline.
 
The comparison shows that the extended candidate set performs best in terms of MAP and P@50 (MAP: 61.42 and P@50: 88.3, cf. Table~\ref{tab:miniComp}). Comparing the MAP results (MAP: 45.56 vs. 35.87) of the two simplified pipelines Gist($S_{t, i}$) and Gist($S_{t}$), indicate the benefit of including information from the image even though it might contain false positive detections. This observation is also confirmed by the result of P@50: 53.71 vs. 49.18).

This study demonstrates that all pipeline steps and the automatic object detection, which allows for an end-to-end approach, increase the performance in the multimedia indexing task. Finally, it confirms the benefit of understanding the gist of multimedia pairs, such as images and texts, in established research domains.    

\section{Conclusion}
To lower the barrier of fostering novel research tasks, such as the one of gist detection and understanding, far beyond the literal meaning of images, we have shown that gist detection performs well on established research questions. We benefit from a pipeline that finds semantic diversity, while also addressing semantic relatedness. Both aspects address the characteristic of the multimedia indexing task: instances can be assigned to several classes, as they convey different semantic meanings, however, at the same time, they share semantics with instances assigned to the same class. 
In nearly 40\% of the general topic and 84\% of the sub-topic cases, the extended gist detection pipeline outperforms the comparison approaches, which implies robustness of our approach. 
Comparing the overall performance, the gist detection pipeline outperforms the Deep Boltzman Machine approach and is comparable to the robust multi-label hashing approach,  where for the latter one no result is given for precision. 
In an end-to-end approach we have demonstrated that multimodal modelling - even if one modality comes from noisy object detectors - performs better than a single modality approach.  



\section*{Appendix A}
\subsection{Seed Node Linking}

\begin{table*}[h!]
	\begin{center}
		\caption[Number of candidates and seed nodes after concept linking (Step 1)]{Number of candidates and seed nodes after the concept linking according to Flickr Tags (text) or image as their origin.
			\label{tab:entityMini}}
			\begin{scriptsize}
		\begin{tabular}{cccccccc}
			\toprule
			
			& \multicolumn{2}{c}{Total} && \multicolumn{2}{c}{Unique} & & Empty Instances\\
			\cmidrule{2-3} \cmidrule{5-6}
			& Candidates & Seed Nodes && Candidates & Seed Nodes & & \\
			\midrule
			Tags & 223,537 & 157,473 && 74,427 & 34,127 & & 2,128 \\
			Images & 514,819 & 487,724 && 976 & 964 & & 166 \\
			\bottomrule
		\end{tabular}
		\end{scriptsize}
	\end{center}
		
\end{table*}

There are no gold standard entity links provided in the dataset and it is not feasible to annotate the entity links for all of the 25,000 pairs. Consequently, we provide a statistical overview on how many of the concept candidate mentions are actually linked to a concept in the knowledge base. 
In Table~\ref{tab:entityMini} the total number of candidates and the finally total number of linked entities are given, which we refer to as seed nodes. 
We provide the numbers separately according to the texts and the images. Furthermore, we study how many unique candidates and seed nodes are found across the dataset. 

The sum of all tags for all image instances (25,000) of the dataset, is over 223,000 tags (cf. Table~\ref{tab:entityMini}), which results in an average of 9 tags per image. 
However, around 2,000 images do not have a tag at all (cf. Table~\ref{tab:entityMini}, Empty Instances). 
For the images around 500,000 image objects are detected and annotated by the Computer Vision API, which is an average of 20 visually recognizable objects per image. 
On 166 images no visually recognizable object is detected, whereas 2 images could not be processed by the API due to an extreme format (e.g., 500px width, 49px height). 

For both types of media - image and text - around half of the candidate concept mentions can be linked to actual concepts in the knowledge base. This results in 157,473 and 487,724 seed nodes for the texts and the images, respectively. 
The statistics about uniqueness across the dataset reveals that the tags are more diverse than the visually recognizable objects: Even though the total number of visually recognizable objects are twice the number of tags, the image objects are from 976 semantic concepts, whereas the tags are from 74,427 semantic concepts.



\clearpage

\bibliographystyle{splncs}
\bibliography{dissRefs}
\end{document}